\documentclass[onecolumn,showpacs,showkeys,preprintnumbers,amsmath,amssymb]{revtex4}

\usepackage{amsmath, amssymb, latexsym, verbatim}
\usepackage{graphicx}% Include figure files
\usepackage{dcolumn}% Align table columns on decimal point
\usepackage{bm}% bold math
\usepackage[dvips]{color}
\usepackage[usenames,dvipsnames]{xcolor}

\newcommand{\nin}{\noindent}
\newcommand{\di}{\partial}
\newcommand{\del}{\nabla}
\newcommand{\deldo}{{\nabla^2}}
\newcommand{\f}{\frac}
\newcommand{\hb}{\hbar}
\newcommand{\ep}{\varepsilon}

\begin{document}

\title{A new way for the extension of quantum theory: Non-Bohmian quantum potentials}

\author{Mahdi Atiq}
 \altaffiliation{Sharif University of Technology, Physics Department, Tehran, Iran}
 \email{mma\_atiq@yahoo.com}

\author{Mozafar Karamian}
% \altaffiliation{Department of Physics and Nuclear Science, Amir Kabir University of Technology, Tehran, Iran}
 \email{karamian@ymail.com}

\author{Mehdi Golshani}
 \altaffiliation{Institutes for Theoretical Physics and Mathematics (IPM), Tehran, Iran}
 \altaffiliation{Department of Physics, Sharif University of Technology, Tehran,
        Iran}
 \email{mehdigolshani@yahoo.com}

\begin{abstract}
\nin Quantum Mechanics is a good example of a successful theory.
Most of atomic phenomena are described well by quantum mechanics
and cases such as Lamb Shift that are not described by quantum
mechanics, are described by quantum electrodynamics. Of course, at
the nuclear level, because of some complications, it is not clear
that we can claim the same confidence. One way of taking these
complications and corrections into account seems to be a
modification of the standard quantum theory. In this paper and its
follow ups we consider a straightforward way of extending quantum
theory. Our method is based on a Bohmian approach. We show that
this approach has the essential ability for extending quantum
theory, and we do this by introducing "non-Bohmian" forms for the
quantum potential.
\end{abstract}

\keywords{Bohmian Mechanics, Quantum Potential, Hamilton-Jacobi,
Variational principle}

\pacs{03.65.Ca, 03.65.Ta, 45.20.Jj, 11.10.Ef}

\maketitle

\emph{Published in Foundations of Physics, 2009}

\section{Introduction} \label{S:Intro}
\nin Quantum mechanics was developed during 1920's and soon became
a successful theory. But in order to explain phenomena like fine
structure of the atomic spectra, e.g., people introduced
relativistic corrections and spin-orbit coupling and finally
Dirac's relativistic formulation of quantum theory superseded all
previous work. But a lot of controversy developed around the
interpretation of the theory. Then, in 1952 David Bohm introduced
another formulation of Quantum theory with a causal interpretation
\cite{Bo52}. Bohm attributed all quantum effects to the so-called
quantum potential that he had introduced in his formulation. But
he kept the basic concepts like wave function, that were used in
the standard formulation of Quantum mechanics. Here we intend to
extend non-relativistic quantum mechanics by using a Bohmian
approach. To begin with, we derive the basic equations of Bohmian
mechanics from a variational principle. Then, we extend the idea
of quantum potential from its simple Bohmian form and try to see
some of the basic results of this new formulation of quantum
theory. The main purpose of this article is to describe a simple
way for the extension of quantum theory by introducing new forms
for the quantum potential. A detailed discussion on the
consequences of these new quantum potentials will come in
following papers.

\section{Bohmian mechanics and the "Quantum Potential" concept} \label{S:BM & QP}
\nin The Schr\"odinger equation is the basic equation of quantum
mechanics:
\begin{equation} \label{E:Shrod Eq}
i\hbar \f{\di}{\di t} \psi=-\f{\hbar^2}{2m} \del^2 \psi+V \psi.
\end{equation}

In the Bohmian interpretation of quantum mechanics, the phase and
amplitude of wave function is meaningful. If we write the wave
function in the polar form
\begin{equation} \label{E:Psi Def} \psi
(x, t) = R (x, t) \, \exp (i S(x, t)/\hbar)
\end{equation}

\nin and substitute it in the complex Schr\"odinger equation
\eqref{E:Shrod Eq}, we obtain the following two real equations:

\begin{equation} \label{E:QHJ Eq}
\f{ {( \del S )}^2 }{2m} + V(x)- \f{\hbar^2}{2m} \f{\del^2 R}{R}+
\f{\di S}{\di t}=0
\end{equation}

\begin{equation} \label{E:Contin Eq}
\f{\di R^2}{\di t}+ \del . ( R^2 \f{\del S}{m} )=0.
\end{equation}

According to Bohm \cite{Bo52}, the Eq. \eqref{E:QHJ Eq} has the
same meaning as the Hamilton-Jacobi equation of classical
mechanics, with an extra term:

\begin{equation} \label{E:QP Def}
Q=-\f{\hbar^2}{2m} \f{\del^2 R}{R}
\end{equation}

\nin which he called it "Quantum Potential" \cite{Bo52}. In Bohm's
view this additional term is responsible for all quantum effects.
Interpreting \eqref{E:QHJ Eq} as a modified Hamilton-Jacobi
equation is meaningful only when we interpret $\del S$ as particle
momentum, and $-{\di S} / {\di t}$ as particle energy (for
conserved systems), i.e.:

\begin{equation} \label{E:p eq. del S}
p=\del S
\end{equation}

\begin{equation} \label{E:E eq. partial t S}
E=-\f{\di S}{\di t}.
\end{equation}

Therefore, we can say that the fundamental assumption of Bohm is
that the phase of wave function 'determines' or 'contains
information about' energy and momentum of particle, i.e. of every
individual member of ensemble of particles. It seems that the lack
of this fundamental assumption in the standard quantum mechanics
is the cause of apparent indeterminism of quantum phenomenon, at
the level of individual members of ensemble. By adding this
assumption, we can talk about the behavior of every particle
alone, along with the behavior of ensemble of them. Without
assuming Eqs. \eqref{E:p eq. del S} and \eqref{E:E eq. partial t
S} it is hard to consider Eq. \eqref{E:QHJ Eq} as a modified
Hamilton-Jacobi equation, and also, without assuming Eq.
\eqref{E:QHJ Eq} as modified Hamilton-Jacobi equation it is hard
to understand that the phase of wave function can contain
information about particle's energy and momentum.

Note that, the Eq. \eqref{E:p eq. del S} is fundamental and
revolutionary assumption only in comparison with the formulation
of orthodox quantum mechanics, not in comparison with the
classical mechanics, because, there is no definite trajectory for
a particle in the formulation of orthodox quantum mechanics, while
in the classical mechanics particle trajectory is a trivial
concept. In contrast, the existence of an extra term, named
quantum potential, in the Hamilton-Jacobi equation in comparison
with the classical mechanics is a fundamental and revolutionary
assumption. Therefore, taking either of the statements $p=\del S$
and $Q=- (\hbar^2 / 2m)(\del^2 R / R) $ as fundamental and
revolutionary depends on wether we compare Bohmian mechanics with
the orthodox quantum mechanics or with the classical mechanics.
The first route provides for the quantum mechanics the possibility
of defining trajectories (something that is absent from quantum
mechanics), and the second one provides for the classical
mechanics the possibility of barrier penetration and other
non-classical behaviors observed in the nature (something that is
absent from classical mechanics). In other words, the Bohmian
mechanics brings quantum and classical mechanics closer together.

Some authors \cite{DGZ96},\cite{Go96} prefer to derive the Eq.
\eqref{E:p eq. del S} from symmetry considerations about the wave
function, but, it seems that arriving at the equation \eqref{E:p
eq. del S} from a Newtonian force law is much simpler than
arriving at it from symmetry considerations about the
Schr\"odinger wave function. Consider the equation

\begin{equation} \label{E:Newton Force Law}
\f{dp}{dt}=\f{\di p}{\di t}+\f{p}{m}.\del p=-\del (V+Q).
\end{equation}

From the identity

\begin{equation} \label{E:Vector Identity}
p.\del p=\del (\f{p^2}{2})+(\del \times p) \times p
\end{equation}

\nin we have

\begin{equation}
\f{\di p}{\di t}+ \del (\f{p^2}{2m})+ (\del \times p) \times
\f{p}{m}=- \del (V+Q)
\end{equation}

\nin or

\begin{equation} \label{E:Curl p identity}
\f{\di}{\di t}(\del \times p)+\del \times ((\del \times p) \times
\f{p}{m})=0.
\end{equation}

The simplest solution of this equation is $p=\del S$  for some
function $S$ (we have extracted the Eqs. \eqref{E:Newton Force
Law}-\eqref{E:Curl p identity} from reference \cite{Gra03}).
Therefore, in the Newtonian framework, we can always take momentum
$p$ as gradient of some function $S$. In this framework, it seems
trivial that we need extra (non-classical) potentials to explain
the non-classical behaviors of particles.

\section{Deriving Bohmian mechanics from a variational principle}
\label{S:BM from variation} \nin We can deduce Bohmian equations
\eqref{E:QHJ Eq} and \eqref{E:Contin Eq} from a variational
principle. This variational principle says that the integral

\begin{equation} \label{E:Variation Integrand}
\int R^2 \left \{ \f{(\del S)^2}{2m}+V-\f{\hbar^2}{2m}\f{\del^2
R}{R}+\f{\di S}{\di t} \right \} d^3 x \, dt
\end{equation}

\nin should be stationary. If we denote the integrand by $g(x, t)$
and treat functions $R$  and $S$ as some fields, then according to
the calculus of variations we have the following equations for
$R$, $S$:

\begin{equation} \label{E:R Eq}
\f{\di g}{\di R}-\di_\alpha \f{\di g}{\di (\di_\alpha R)} +
\di_\alpha \di_\beta \f{\di g}{\di (\di_\alpha \di_\beta R)}=0
\end{equation}

\begin{equation} \label{E:S Eq}
\di_\alpha \f{\di g}{\di (\di_\alpha S)}=0.
\end{equation}

The indices $\alpha$, $\beta$  refer to space and time
coordinates. From \eqref{E:R Eq} we obtain

\begin{equation} \label{E:R Eq Expand}
2R \left \{ \f{(\del S)^2}{2m}+V-\f{\hbar^2}{2m}\f{\del^2
R}{R}+\f{\di S}{\di t} \right \} + R^2 \left \{ \f{\hbar^2}{2m}
\f{\del^2 R}{R^2} \right \} + \del^2 \left \{ R^2 (-
\f{\hbar^2}{2m} \f{1}{R}) \right \}=0
\end{equation}

\nin which reduces to the modified Hamilton-Jacobi equation
\eqref{E:QHJ Eq}. Similarly from \eqref{E:S Eq} we have

\begin{equation}
\del . (R^2 \f{\del S}{m})+\f{\di}{\di t}R^2=0
\end{equation}

\nin which is the continuity equation \eqref{E:Contin Eq}.
Apparently this is not a remarkable fact, but there is a crucial
and subtle point here. The crucial point is "the form of
functional dependence of the quantum potential with respect to
$R$". The vanishing of the expression

\begin{equation}
R^2 \left \{ \f{\hbar^2}{2m} \f{\del^2 R}{R^2} \right \} + \del^2
\left \{ R^2 (- \f{\hbar^2}{2m} \f{1}{R}) \right \}
\end{equation}

\nin in \eqref{E:R Eq Expand} is due to the special form of
functional dependence of $Q$ with respect to $R$. If we need to
obtain a Hamilton-Jacobi equation modified with a quantum
potential, $Q[R]$ can not take every arbitrary form. The Eq.
\eqref{E:R Eq} restricts the acceptable forms of $Q[R]$. If we
choose another form, $Q = 0$, i.e. we replace the Bohmian quantum
potential $-(\hbar^2 / 2m)(\del^2 R / R)$ in the integral
\eqref{E:Variation Integrand} by a zero constant, then from
\eqref{E:R Eq} we obtain

\begin{equation}
\f{(\del S)^2}{2m}+V+\f{\di S}{\di t}=0
\end{equation}

\nin which is exactly the classical Hamilton-Jacobi equation.
Therefore, by choosing proper and acceptable forms for the quantum
potential we can obtain the classical or the quantum
Hamilton-Jacobi equation. Thus it seems natural to ask: "Can there
be other forms of quantum potential?". As we show in this paper
the answer is positive. In the next section we deduce other forms
("non-Bohmian" forms) of quantum potential.

\section{Other possible forms of quantum potential}
\label{S:Other QPs} \nin In this section for simplicity we write
derivatives as subscripts:

\begin{equation}
R_\alpha \equiv \di_\alpha R, \,\, R_{\alpha \beta} \equiv
\di_\alpha \di_\beta R, \,\, \cdots; \;\; \alpha, \beta \in \{ 0,
1, 2, 3 \}.
\end{equation}

\nin In this notation the Bohmian quantum potential is written as:

\begin{equation} \label{E:BQP}
Q=-\f{\hbar^2}{2m}R^{-1} \delta^{ij} R_{ij}; \;\; i,j \in \{ 1, 2,
3 \}
\end{equation}

\nin in which $\delta^{ij}$  is the Kronecker delta tensor.

For deriving other possible forms of quantum potential we replace
"Bohmian" quantum potential $-(\hbar^2 / 2m)(\del^2 R / R)$  by a
general unknown function $Q$ that is some function of $R$ and its
derivatives i.e.:

\begin{equation}
Q=Q(R, \, , R_\alpha, \,  R_{\alpha \beta}, \, R_{\alpha \beta
\sigma}, \, \cdots ); \;\;  \alpha, \beta, \sigma, \cdots \in \{
0, 1, 2, 3 \}.
\end{equation}

Now, the variational principle is written as

\begin{equation}\label{E:Var Principle}
\delta \int g d^3 x \, dt = \delta \int R^2 \left \{ \f{(\del
S)^2}{2m}+V+Q+\f{\di S}{\di t} \right \} d^3 x \, dt=0.
\end{equation}

\nin in which we have denoted the integrand by $g$. Similar to
Eqs. \eqref{E:R Eq}, \eqref{E:S Eq} in the previous section, we
obtain from variational calculus

\begin{equation} \label{E:R Eq2}
\f{\di g}{\di R}- \di_\alpha \f{\di g}{\di R_\alpha}+ \di_\alpha
\di_\beta \f{\di g}{\di R_{\alpha \beta}}- \di_\alpha \di_\beta
\di_\sigma \f{\di g}{\di R_{\alpha \beta \sigma}}+ \cdots=0
\end{equation}

\begin{equation} \label{E:S Eq2}
\di_\alpha \f{\di g}{\di S_\alpha}=0.
\end{equation}

\nin The Eq. \eqref{E:S Eq2} reduces to the continuity equation.
But, from Eq \eqref{E:R Eq2} we get

\begin{equation}
2R \left \{ \f{(\del S)^2}{2m}+V+Q+\f{\di S}{\di t} \right \}+ R^2
\f{\di Q}{\di R}-\di_\alpha \big( R^2 \f{\di Q}{\di R_\alpha}
\big)+ \di_\alpha \di_\beta \big( R^2 \f{\di Q}{\di R_{\alpha
\beta}} \big)-\di_\alpha \di_\beta \di_\sigma \big( R^2 \f{\di
Q}{\di R_{\alpha \beta \sigma}} \big)+\cdots=0.
\end{equation}

\nin If we expect to have modified Hamilton-Jacobi equation in the
form

\begin{equation}\label{E:General QHJ Eq}
    \f{(\del S)^2}{2m} + V + Q + \f{\di S}{\di t} = 0
\end{equation}

\nin then, we obtain an important condition that the general
quantum potential must fulfil:

\begin{equation} \label{E:QP Contidtion x, t}
R^2 \f{\di Q}{\di R}-\di_\alpha \big( R^2 \f{\di Q}{\di R_\alpha}
\big)+ \di_\alpha \di_\beta \big( R^2 \f{\di Q}{\di R_{\alpha
\beta}} \big)-\di_\alpha \di_\beta \di_\sigma \big( R^2 \f{\di
Q}{\di R_{\alpha \beta \sigma}} \big)+\cdots=0; \;\; \alpha,
\beta, \sigma, \cdots \in \{ 0, 1, 2, 3 \}.
\end{equation}

Now, as in Eq. \eqref{E:BQP} we assume that general quantum
potential does not depend on time derivatives. Then, the Eq.
\eqref{E:QP Contidtion x, t} reduces to

\begin{equation} \label{E:QP Contidtion x}
R^2 \f{\di Q}{\di R}-\di_i \big( R^2 \f{\di Q}{\di R_i} \big)+
\di_i \di_j \big( R^2 \f{\di Q}{\di R_{i j}} \big)-\di_i \di_j
\di_k \big( R^2 \f{\di Q}{\di R_{i j k}} \big)+\cdots=0; \;\; i,
j, k, \cdots \in \{1, 2, 3 \}.
\end{equation}

Note that the solutions $Q[R]$ of this equation must be such that
for every arbitrary function $R$, the equation be satisfied. The
simplest solution is

\begin{equation} \label{E:Q=const}
    Q=A=const
\end{equation}

\nin which is a trivial quantum potential. The next simple form is

\begin{equation}\label{E:Q=Bohmian form}
    Q = A R^{-1} \delta^{ij} R_{ij}
\end{equation}

\nin which is the same as the Bohmian quantum potential. By a
little effort we find all forms of $Q$ in the forms

\begin{equation}\label{E:Q Forms}
    Q_0 = A_0, \,\, Q_2 = A_2 R^{-1} \delta^{ij} R_{ij}, \,\, Q_4 = A_4
    R^{-1}\delta^{ij} \delta^{kl} R_{ijkl}, \,\, \cdots
\end{equation}

\nin which contain only even-order derivatives of $R$, and satisfy
the Eq. \eqref{E:QP Contidtion x}. We did not find an expression
for $Q$ which contains odd-order derivatives of $R$ and which
satisfies the condition \eqref{E:QP Contidtion x}. In fact, we
looked for simple forms for $Q$ and we didn't find any solution.
There may be some complicated forms of $Q$ which contains
odd-order derivatives of $R$. But, we must pay attention that $Q$
is a scalar and the derivatives of $R$ must appear in the form of
operator $\del$. \footnote{Indeed, if we restrict $Q[R]$ to take
only derivatives up to second order, we can prove that the only
possible forms of $Q[R]$ are \eqref{E:Q=const} and
\eqref{E:Q=Bohmian form} (to be published). The extension of the
proof to cover the forms \eqref{E:Q Forms} is under
consideration.}

Therefore, the possible forms of quantum potential are

\begin{equation}\label{E:QP forms}
    Q_0 = A_0, \,\, Q_2 = A_2 R^{-1} \deldo R, \,\, Q_4 = A_4
    R^{-1} \deldo (\deldo R), \,\, \cdots
\end{equation}

\nin in which $A_0$, $A_2$, $A_4$, $\cdots$ are some constants.

\section{On the meaning of new quantum potentials}
\nin In the previous section we obtained new forms for quantum
potential. Now, we take an assumption that the quantum phenomena
arises from the combination of all quantum potentials, i.e. all
quantum potentials contribute to the creation of quantum effects.
Therefore, we can define the \emph{complete} quantum potential as

\begin{equation}\label{E:Complete QP}
    Q = R^{-1} (A_0 R + A_2 \deldo R + A_4 \deldo (\deldo R) + \cdots
    ).
\end{equation}

In this view, the Bohmian quantum potential $Q_2$ is only a first
approximation to the complete quantum potential. The first term
$Q_0$ is a constant and therefore has no role in the dynamics of
particles. However, due to its dimension one can interpret it as
the rest energy of the particle. But, what can we say about the
other terms?

We know that the Schr\"odinger equation can describe all the
non-relativistic quantum phenomena at the atomic level. Therefore,
if these new quantum potentials have physical reality, they must
carry some corrections at the atomic level. The mathematical form
of the terms in \eqref{E:Complete QP} can help us in interpreting
them. We know from the formalism of quantum mechanics that the
operator $\del$ in $Q_2$ is representative of momentum operator.
It is natural to extend this assumption for all terms. Indeed, we
can write the Bohmian term as

\begin{equation}
    Q_2 = \f{1}{2} \ep_0 \bigg( \f{-i \hb}{m c} \bigg)^2 R^{-1} \deldo R,
    \,\, \ep_0 = m c^2
\end{equation}

\nin therefore, we suggest the following form for all terms

\begin{equation}\label{E:}
    Q_{2n} = a_{2n} \ep_0 \bigg( \f{-i \hb}{m c}
    \bigg)^{2n} R^{-1} \del^{2n} R,  \,\, \ep_0 = m c^2, \;\; n \in \{0, 1, 2, 3, \cdots \}
\end{equation}

\nin in which $\varepsilon_0$ is the rest energy, the coefficients
$a_{2n}$ are some dimensionless constants, and we used definition

\begin{equation}
    \del^{2n} = \underset{n \,\, times}{ \underbrace{\deldo \deldo \cdots \deldo}
    }.
\end{equation}

\nin Consequently, the complete quantum potential becomes

\begin{equation}\label{E:}
    Q = \ep_0 R^{-1} \sum_{n=0}^\infty a_{2n} \bigg( \f{-i \hb \del}{m c}
    \bigg)^{2n}R.
\end{equation}

Among the coefficients $a_{2n}$ we only know $a_2 = 1/2 (n=1)$.
Rewriting this suggested form of $Q$ as

\begin{equation}\label{E:}
    Q = \ep_0 R^{-1} \sum_{n=0}^\infty a_{2n} \bigg( \f{\hat{p} c}{\ep_0}
    \bigg)^{2n}R
\end{equation}

\nin we observe that by choosing proper values for $a_{2n}$:

\begin{equation}
    a_{2n} = (-1)^{n+1} \f{1}{2 n - 1} \f{1}{2^{2n}}
    \f{(2n)!}{(n!)^2}, \; n=0, 1, 2, 3, \cdots
\end{equation}

\nin the operator

\begin{equation}\label{E:QPs Op.}
    \ep_0 \sum_{n=0}^\infty a_{2n} \bigg( \f{\hat{p} c}{\ep_0}
    \bigg)^{2n}
\end{equation}

\nin becomes exactly the relativistic energy operator

\begin{equation}\label{E:Rel. Energy Op.}
   \hat{\ep} \equiv \ep_0 \left[ 1 + \bigg( \f{\hat{p} c}{\ep_0} \bigg)^2
   \right]^{1/2}.
\end{equation}

This is an important point because, as we explain in the next
section, we can get corrections to be exactly same as relativistic
corrections to the energy, but now as effect of non-Bohmian
quantum potentials. In other words, by choosing proper values for
the coefficients of new quantum potentials, we can have
corrections equal to relativistic corrections even in the
framework of non-relativistic quantum mechanics, and without
appealing to relativistic \emph{concepts}. Identifying the
operator \eqref{E:QPs Op.} with \eqref{E:Rel. Energy Op.}, we have
for $Q_4$

\begin{equation}\label{E:Q4}
    Q_4 = - \f{1}{8} \f{\hb^4}{m^3 c^2} \f{\deldo(\deldo R)}{R}.
\end{equation}

In the next section, we describe the effect of this term on the
energy levels of atoms.

\section{Relativistic corrections to energy arise from non-Bohmian quantum potentials}
\nin First, we note that for stationary states the time integral
in the Eq. \eqref{E:Var Principle} becomes a coefficient and
therefore the variational principle reduces to

\begin{equation}\label{E:Var Principle for Stationary}
\delta \int g \, d^3 x =\\
\delta \int R^2 \left \{ \f{(\del S)^2}{2m}+V+Q - E \right \} d^3
x = 0.
\end{equation}

\nin Now, suppose that we have for first approximation

\begin{equation}
    Q = Q_0 + Q_2
\end{equation}

\nin and a set of $R_0$, $S_0$ and $E_0$ satisfy the equation
\eqref{E:Var Principle for Stationary}. Then, we have

\begin{equation}\label{E:int g0}
\int g_0 \, d^3 x =\\
\int R_0^2 \left \{ \f{(\del S_0)^2}{2m}+ V + Q_0 + Q_2 - E_0
\right \} d^3 x.
\end{equation}

Suppose that the problem with $Q_4$ yield small shifts in energy
and that $R_0$, $S_0$ remain approximately as before, we can write
for $Q_4$

\begin{equation}\label{E:Q4(R0)}
    Q_4 \simeq - \f{1}{8} \f{\hb^4}{m^3 c^2} \f{\deldo(\deldo R_0)}{R_0}.
\end{equation}

\nin and we have

\begin{equation}\label{E:int g}
\int g \, d^3 x =\\
\int R_0^2 \left \{ \f{(\del S_0)^2}{2m}+ V + Q_0 + Q_2 + Q_4 -
(E_0 + \Delta E) \right \} d^3 x.
\end{equation}

\nin Assuming

\begin{equation}\label{E:Shift in Energy 0}
    \Delta E = \int R_0^2 Q_4 d^3 x
\end{equation}

\nin the integral \eqref{E:int g} reduces to \eqref{E:int g0} and
therefore satisfies the variational principle \eqref{E:Var
Principle for Stationary}. If we rewrite the Eq. \eqref{E:Shift in
Energy 0} as

\begin{equation}\label{E:Shift in Energy}
    \Delta E = \int R_0^2 \left \{ - \f{1}{8} \f{\hb^4}{m^3 c^2} \f{\deldo(\deldo R_0)}{R_0} \right \} d^3 x
\end{equation}

\nin and compare it with the 'relativistic correction'(in the
usual quantum mechanics)

\begin{equation}\label{E:Rel. Correction}
    \Delta E_{rel} = \int \psi_0^\star \left \{ - \f{(\hat{p}^2)^2}{8 m^3 c^2} \right \} \psi_0 d^3 x
\end{equation}

\nin then we observe that if $S_0 = const$, the Eqs. \eqref{E:Rel.
Correction} and \eqref{E:Shift in Energy} becomes identical and
the value of $\Delta E$ becomes exactly the relativistic
correction shift in energy.

Now, we consider the energy eigenstates of central potentials.
These states are written as

\begin{equation}\label{E:Usual states}
    \psi = R_{nl}(r) P_l^m(\cos \theta) e^{im \phi}.
\end{equation}

But, there is a degeneracy due to the spherical symmetry of the
potential in quantum number $m$. Due to this degeneracy, we can
write the state functions as

\begin{equation}
\begin{aligned}
    &\psi = R_{nl}(r) P_l^m(\cos \theta) \cos m \phi\\
    &\psi = R_{nl}(r) P_l^m(\cos \theta) \sin m \phi
\end{aligned}
\end{equation}

\nin in place of \eqref{E:Usual states}. Therefore, it is possible
to have $S_0 = const$ for all energy eigenstates of the central
potential. Consequently, we can assume that electrons in all
eigenstates of energy are at rest and the 'relativistic
correction' comes from non-Bohmian quantum potential $Q_4$.
Indeed, by considering higher order terms in the quantum potential
\eqref{E:Complete QP} we can get the relativistic corrections to
every order.

Note that, the fact that $Q_4$ is proportional to $R^{-1}
\deldo(\deldo R)$ or $R^{-1} (\hat{p}^2)^2 R$ is a consequence of
generalization of quantum potential in the framework of
non-relativistic Bohmian mechanics. Choosing a proper value for
the coefficient $A_4$ in $Q_4$, this term can yield a correction
to energy that is exactly equal to what we call 'the relativistic
correction'.

\section{Schr\"odinger equation as an approximation}
\nin We know from ordinary Bohmian mechanics that the set of
equations \eqref{E:QHJ Eq} and \eqref{E:Contin Eq} for Bohmian
quantum potential is equivalent to the Schr\"odinger equation
\eqref{E:Shrod Eq}. But, by including non-Bohmian quantum
potentials in the modified Hamilton-Jacobi equation, one can not
obtain a linear Schr\"odinger-like equation for the wave function
$\psi = R \exp(iS/ \hb)$. Therefore, the Schr\"odinger equation
remains as an approximation. Indeed, one can combine the
continuity and modified Hamilton-Jacobi equations in the presence
of complete quantum potential to obtain the equation

\begin{equation}\label{E:New Shrod. Eq.}
\begin{aligned}
    i \hb \f{\di \psi}{\di t} &= \left( - \f{\hb^2}{2 m} \deldo + V + Q - Q_2 \right)
    \psi \\
    &=\left( - \f{\hb^2}{2 m} \deldo + V + Q_0 + Q_4 + Q_6 + \cdots \right)
    \psi.
\end{aligned}
\end{equation}

\nin We observe that, only when $Q \simeq Q_0 + Q_2$, we have a
linear equation. But, we remember that the Shr\"odinger equation
describes well all non-relativistic quantum phenomena at the
atomic level. Therefore, the effects of non-Bohmian quantum
potentials $Q_4, Q_6, \cdots$ at the atomic level in comparison
with the effect of Bohmian quantum potential $Q_2$ must be small
and consequently the Schr\"odinger equation and its linearity
remains a good approximation. This requirement must be considered
in the interpretation of non-Bohmian quantum potentials and the
suggestion for their coefficients. This is the case for our
suggested interpretation of new quantum potentials. For example
consider a particle in a one-dimensional box with the length $L$.
Assuming eigenfunction $\sin (\tau \pi x / L)$ we have

\begin{equation}
    \f{Q_{2(n+1)}}{Q_{2n}} = \f{a_{2(n+1)}}{a_{2n}} \bigg( \f{\tau \pi \hb}{m c L}
    \bigg)^2 = \f{a_{2(n+1)}}{a_{2n}} \tau^2
    \bigg( \f{\lambda_c}{2L} \bigg)^2
\end{equation}

\nin in which $\tau$ is quantum number of energy level and
$\lambda_c$ is Compton wavelength. For an electron in the atomic
dimensions($L \sim \mathring{A}$), the proportion $( \lambda_c /
2L )^2$ is of the order of $10^{-4}$, but for a proton in nuclear
dimensions($L \sim 10^{-5} \mathring{A}$), it is of the order of
$10^{-1}$. Therefore, the effects of new quantum potentials at the
atomic level are in the form of small perturbations but at the
nuclear level these effects are remarkable. Note that, for a
proton too in the \emph{atomic} dimensions($\sim \mathring{A}$),
these effects are small. Therefore, we conclude that the
Schr\"odinger equation is adequate at the atomic level but at the
nuclear level this equation is not necessarily adequate.

We must mention that Bohm itself foresaw the possibility of
including nonlinear terms to the Schr\"odinger equation which are
large only for processes involving small distances \cite{Bo52}.
Our suggestion for the coefficients of new quantum potentials has
this merit that the effects of higher order quantum potentials
become large only for the dimensions which are smaller than the
atomic distances. Therefore, the linearity of Schr\"odinger
equation and the superposition of states can be preserved only at
the atomic level or larger dimensions. If our interpretation of
new quantum potentials is correct, we can expect some highly
nonlinear effects at the nuclear or smaller dimensions. The lack
of superposition affects the global structure of quantum
mechanics. For example, we expect some serious effects on the
EPR-Bell type nonlocal correlations at the nuclear level. However,
some properties of Bohmian quantum potential such as its not
falling with distance and its dependence on the form of wave
function are preserved. Bohm described the apparent jumps and
discontinuity of atomic processes by the behavior of quantum
potential. According to Bohm, due to the form of dependence of
(Bohmian) quantum potential on the wave function, even when the
amplitude of wave function is very small, we can expect some large
energy exchanges at very short times. This property is preserved
for new quantum potentials, due to their dependence on $R$.
Therefore, the existence of sharp energy levels and the jumping
between them at the nuclear level, which have experimental
evidence, are permitted by non-Bohmian quantum potentials. In a
following paper, we shall discuss in details the main effects of
new quantum potentials, such as nonlinearity, and the effects on
the global structure of quantum theory and phenomena such as
EPR-Bell nonlocality.

\section{Conclusion}
\nin In this article we obtained new forms for quantum potential
and assumed that quantum effects are due to the combination of all
quantum potentials. Due to mathematical form of the series of
complete quantum potential, we suggested that the coefficients of
quantum potentials can be chosen so that they yield corrections
equal to the relativistic corrections to energy. An important
point about our approach is that the mathematical form of the new
quantum potentials (i.e. the functional dependence of them with
respect to $R$) is a consequence of generalization of quantum
potential in the framework of non-relativistic Bohmian mechanics,
without any reference to relativistic concepts. We observed that
choosing proper values for the coefficients makes the effects of
non-Bohmian quantum potentials mathematically equivalent to what
we get from relativistic effects. The importance of this
observation is that to obtain the relativistic corrections to
energy there is no necessity to start from relativistic
considerations. For a better understanding of this subject
consider classical electrodynamics. The classical electrodynamics
was completed before relativity and was a motivation for it. This
means that for the derivation of the "correct" (i.e. relativistic)
theory of electrodynamics there is no need to start from comparing
one observer to another. The Maxwell's equations are all
consequences of experiments that are done in the reference frames
which are at rest relative to the Earth. But these experiments
were sufficient to yield the correct equations of electrodynamics.
Our considerations about new quantum potentials may be similar to
this situation. There may be a possibility to obtain "correct"
mechanical equations without using relativistic "concepts" and
without comparing observers with each other, and the consideration
of non-Bohmian quantum potentials may be only a first step for
realizing this demand.

\end{document}